\documentclass{jfm}
\usepackage{graphicx}
\usepackage{epstopdf, epsfig}
\usepackage{xcolor}

\shorttitle{Crown formation}
\shortauthor{Y. Saade, M. Jalaal, A. Prosperetti and D. Lohse}

\title{Crown formation from a cavitating bubble close to a free surface}

\author{Youssef Saade\aff{1}
  \corresp{\email{y.saade@utwente.nl}},
  Maziyar Jalaal\aff{2}
  \corresp{\email{m.jalaal@uva.nl}},
  Andrea Prosperetti\aff{1,3}
  \corresp{\email{aprosper@central.uh.edu}},
 \and Detlef Lohse\aff{1,4}\corresp{\email{d.lohse@utwente.nl}}}

\affiliation{\aff{1}Physics of Fluids Group, Max Planck Centre for Complex Fluid Dynamics, MESA+ Institute\\ and J.M. Burgers Centre for Fluid Dynamics, University of Twente,\\P.O. Box 217, 7500 AE Enschede, The Netherlands
\aff{2}Van der Waals-Zeeman Institute, University of Amsterdam, Science Park 904, Amsterdam, The Netherlands
\aff{3}Department of Mechanical Engineering, University of Houston, TX 77204-4006, USA
\aff{4}Max Planck Institute for Dynamics and Self-Organisation, Am Fassberg 17, 37077 Göttingen, Germany}

\begin{document}

\maketitle

\begin{abstract}
	A rapidly growing bubble close to a free surface induces jetting: a central jet protruding outwards and a crown surrounding it at later stages. While the formation mechanism of the central jet is known and documented, that of the crown remains unsettled. We perform axisymmetric simulations of the problem using the free software program $\textsc{basilisk}$, where a finite-volume compressible solver has been implemented, that uses a geometric Volume-of-Fluid method (VoF) for the tracking of the interface. We show that the mechanism of crown formation is a combination of a pressure distortion over the curved interface, inducing flow focusing, and of a flow reversal, caused by the second expansion of the toroidal bubble that drives the crown. The work culminates in a parametric study with the Weber number, the Reynolds number, the pressure ratio and the dimensionless bubble distance to the free surface as control parameters. Their effects on both the central jet and the crown are explored. For high Weber numbers, we observe the formation of weaker ``secondary crowns", highly correlated with the third oscillation cycle of the bubble.
\end{abstract}


\section{Introduction}
\begin{figure}
	\centerline{\includegraphics[width=\textwidth]{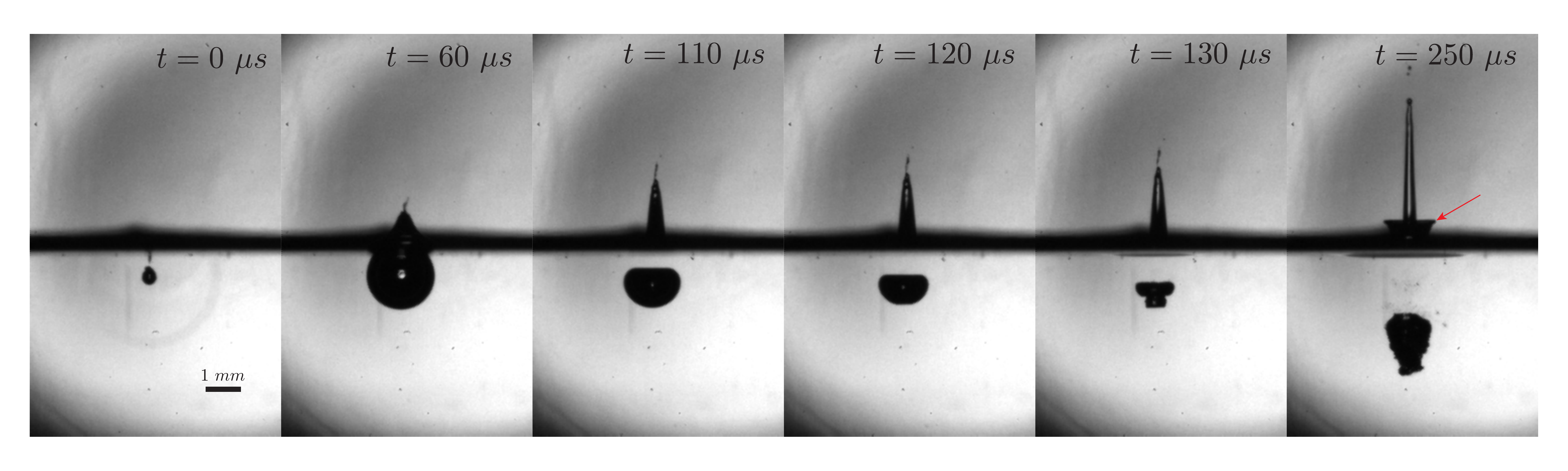}}
	\caption{A pulsed laser is focused in water, close to the free surface. At $t = 0 \ \mu s$, a highly pressurized bubble is generated, emitting a shock wave. The bubble then aspherically expands into an egg-shape, with the tip pointing upwards ($t = 60 \mu s$). This leads to an asymmetric collapse, where the bubble is pierced by an inner jet ($t = 110 \ \mu s$) which then breaks it into a toroid ($t = 120 \ \mu s$). At the same time, a jet is protruding upwards. The toroidal bubble further breaks into two toroids ($t = 130 \ \mu s$), and expands again while travelling downwards ($t = 250 \ \mu s$). Then the formation of an axisymmetric crown around the central jet is observed, see the arrow in the last snapshot. The experimental snapshots were provided by Dave Kemper (Physics of Fluids lab, University of Twente).}
	\label{fig:expseq}
\end{figure}

Free surface jetting is a widely observed phenomenon in nature with several practical applications. Of particular relevance to this paper is its occurrence in the laser-induced forward transfer (LIFT) process, a digital printing technique where parts of a donor film are transferred to a receiving substrate \citep{37,38,1,16,3,17,chahine,thoroddsen,pain2012jets,bempedelis2021numerical}. Figure \ref{fig:expseq} shows a typical sequence of events observed in a laboratory experiment related to the LIFT process: due to optical/thermal breakdown via focusing of a pulsed laser, a bubble nucleates, expands and collapses while a jet is ejected upwards from the free surface. In addition, an unexpected and rather remarkable phenomenon, namely the formation of an axisymmetric crown around the rim of the jet, is observed about $100 \ \mu s$ after the ejection of the central jet. While it is clear that the central jet is caused by the rapidly expanding bubble, the mechanism responsible for the crown observed in this as well as in other circumstances is still controversial.  

Several examples of jets associated with cavitation bubbles can be found in the literature. \citet{20} and \citet{15} studied laser-induced bubbles inside a cylindrical water jet close to the free surface, which may be considered as a variation of the experiment shown in figure \ref{fig:expseq} in which the free surface is initially plane. Instead of an axisymmetric crown, the authors observe two in-plane micro jets. They speculated that this might be the effect of a shock wave impinging on the curved water interface. \citet{13} studied the jet formation inside a capillary tube due to laser-induced cavitation. The authors highlighted the compressibility effects and argued that the jet is likely due to the impingement of the shock wave on the initially curved meniscus. \citet{12} studied cavitation inside a droplet, leading to a corrugated free surface due to the Rayleigh-Taylor instability. The authors attribute the subsequent jetting to pressure impulses that focus the flow on suitably curved interfaces. \citet{14} performed tube impact tests as in \citet{11}, but for stronger impact. They observed cavitation in the tube and then the formation of crowns around the central jet. They argue that the interaction of expansion and compression waves with the tube wall and the curved interface results in the crown formation.

In LIFT, the focus on liquid compressibility as a determining factor in the formation of the secondary jet, i.e., the crown, is put into question by the work of \citet{47} and \citet{48}, who both used an incompressible Boundary Element Method (BEM) to study similar configurations. Due to the  challenges that the BEM encounters in simulating the bubble after it breaks into a toroid, the former authors simply removed the toroidal bubble. The latter authors  studied the interaction of two, vertically adjacent, oscillating bubbles close to a free surface but, unlike \citet{47}, they maintained the bubble even after it became a toroid and they observed a trace of a crown.

Other numerical studies dispensing with the assumption of liquid incompressibility exist. \citet{18} performed Volume-of-Fluid (VoF) simulations with the aim of replicating their own experimental results. They achieved an overall good agreement, both qualitatively and quantitatively. Although their numerical results exhibit a crown, they do not discuss it nor make any comment on the mechanism of its formation or on how it depends on the control parameters. \citet{19} also used a VoF method to simulate an oscillating bubble close to a free surface, also achieving a rather good agreement with their experiments. In addition, they performed a parametric study, varying the initial bubble-to-interface distance, observing bursting behaviour in some instances. On the subject of crown formation, they write ``when the toroidal bubble starts to rebound, the pushing effect from the rebounding bubble, similar to the effect from the initial expanding bubble, tends to induce the upward deformation of the free surface again, which is the cause of the secondary spike". They add ``since [\citet{48}] ignored the compressibility of the water, acoustic emissions were absent in their simulation. Therefore, we argue that the acoustic emissions are not important in the formation of the crown spike". Beyond these statements, however, they did not provide any quantitative evidence of the role of compressibility.

In the present work, we perform Direct Numerical Simulations (DNS) of a cavitation bubble in the vicinity of a free surface, focusing in particular, unlike the work of \citet{18} and \citet{19}, on the formation of the crown, and studying to what extent liquid compressibility contributes to the phenomenon. Furthermore, we perform a parametric study, analysing the effect of surface tension, viscosity, pressure ratio and bubble-to-interface initial distance on the dynamics.

\begin{figure}
	\centerline{\includegraphics[width=0.8\textwidth]{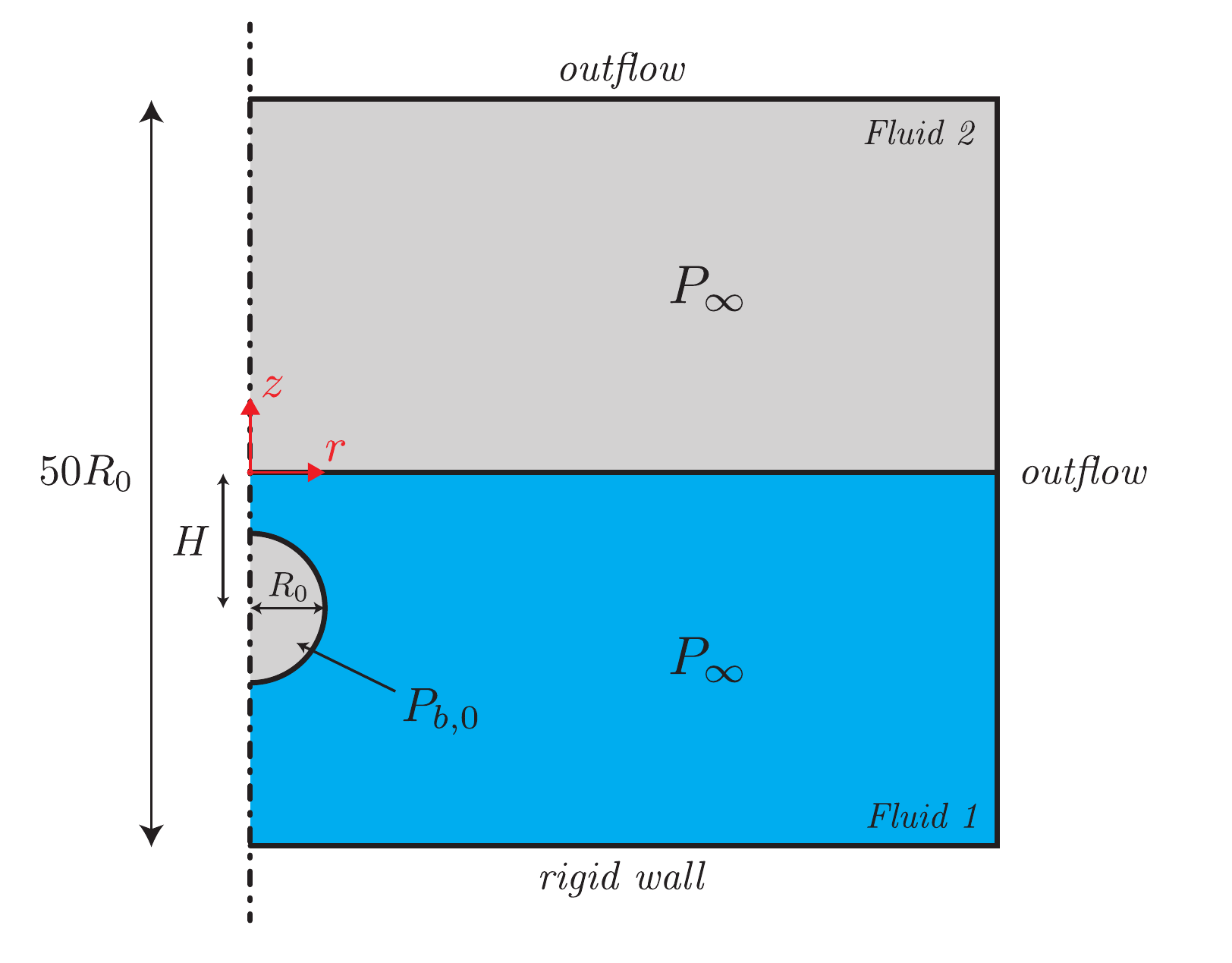}}
	\caption{The numerical setup for the study of a cavitation bubble in the vicinity of a free surface. The blue depicts the heavier fluid (e.g. water) while grey represents the lighter one (e.g. air), respectively referred to as fluids $1$ and $2$ and having an initial density ratio of $\rho_1/\rho_2 = 1000$. The shown dimensions are not to scale.}
	\label{fig:numset}
\end{figure}

The paper is organised as follows. In \S\ref{sec:num}, we present the governing equations, the numerical method, the numerical setup, and the relevant dimensionless numbers. In \S\ref{sec:crown}, we reiterate in more detail the above sketched dynamics of the phenomenon, focusing on the origin of crown formation. In \S\ref{sec:param}, we present the parametric study. Finally, we summarize our findings in \S\ref{sec:conclusion}.
 
\section{Geometry, governing equations, control parameters, and numerical scheme}\label{sec:num}
\subsection{Setup}\label{subsec:set}
The numerical setup is shown in figure \ref{fig:numset}. A bubble of initial radius $R_0$ is placed in a liquid at a distance $H$ between its centre and the liquid's interface. The bubble has an initial internal pressure $P_{b,0}$ larger than the surrounding pressure $P_\infty$. The bottom boundary has the conditions of a rigid wall, i.e. $\boldsymbol{u}=0$, and a zero pressure gradient in the normal direction ($\partial p/\partial z = 0$). The top and right boundaries have outflow conditions where we impose the pressure as $p = P_\infty$, zero normal velocity gradients and vanishing shear stresses (top: $\partial v_r/\partial z = 0$, $\partial v_z/\partial z = 0$, right: $\partial v_z/\partial r = 0$, $\partial v_r/\partial r = 0$). Gravity is neglected ($g = 0$) since we are studying millimetric bubbles with a short lifetime where buoyancy effects are small \citep{26}, as opposed to UNDEX problems where the bubble sizes are much larger and buoyant forces become dominant \citep{jfm-zhang-2015}. Hence, we use the constant atmospheric pressure $P_\infty$ at the right boundary instead of the hydrostatic pressure. All the simulations are performed in axisymmetric configuration. The cylindrical domain is $50$ times the initial radius of the bubble, both in radius and height.  Its relatively large size is chosen so as to diminish the effect of boundaries on the problem, especially the reflection of pressure waves and their interaction with the physical process \citep{21}.

\subsection{Governing equations}\label{subsec:eqs}
The governing equations of this two-phase flow (subscript $i$ denoting the different phases, equal to 1 in the liquid, and to 2 in the gas), in their conservative form,  are written as follows:
\begin{equation}
	\frac{\p\rho_i}{\p t} + \bnabla\bcdot\left(\rho_i\boldsymbol{u}_i\right) = 0,
	\label{eq:continuity}
\end{equation}
\begin{equation}
	\frac{\p\rho_i\boldsymbol{u}_i}{\p t} + \bnabla\bcdot\left(\rho_i\boldsymbol{u}_i\boldsymbol{u}_i\right) = \bnabla\bcdot\mathsfbi{T}_i + \sigma\kappa\delta_s\boldsymbol{n},
	\label{eq:momentum}
\end{equation}
which reflect conservation of mass and momentum, respectively. In the equations above,  $\rho$ is the density and $\boldsymbol{u}$ is the velocity vector. The last term in equation (\ref{eq:momentum}) deals with the capillary forces and is written in a discrete form, where $\delta_s$ is a characteristic function defined as $1$ in cells containing an interface, and otherwise $0$; $\sigma$ is the surface tension coefficient, $\kappa$ the interface curvature and  $\boldsymbol{n}$ the normal to the interface. The stress tensor is
\begin{equation}
	\mathsfbi{T}_i = -\left(p_i + \frac{2}{3}\mu_i\bnabla\bcdot\boldsymbol{u}_i\right)\mathsfbi{I} + \mu_i\left(\bnabla\boldsymbol{u}_i + \bnabla\boldsymbol{u}_i^T\right),
	\label{eq:stresstensor}
\end{equation}
where $p$ is the pressure and $\mu$ is the shear viscosity. Note that, following Stokes' hypothesis, the bulk viscosity is neglected \citep{tcps-stokes-1845}. In the present work, mass transfer and thermal diffusion effects are neglected. The total energy equation is then written as:
\begin{equation}
	\frac{\p}{\p t}\left[\rho_i\left(e_i + \frac{1}{2}\boldsymbol{u}_i^2\right)\right] + \bnabla\bcdot\left[\rho_i\left(e_i + \frac{1}{2}\boldsymbol{u}_i^2\right)\boldsymbol{u}_i\right] = \bnabla\bcdot\left(\mathsfbi{T}_i\bcdot\boldsymbol{u}_i\right),
	\label{eq:totalenergy}
\end{equation}
where $e$ is the internal energy of the fluid. The system is closed by an equation of state (EOS) relating the different thermodynamic properties. We use the stiffened-gas EOS, written in the Mie-Grüneisen form \citep{35}:
\begin{equation}
	\rho_i e_i = \frac{p_i + \Gamma_i\Pi_i}{\Gamma_i - 1},
\end{equation}
from which the speed of sound follows as:
\begin{equation}
	c_i = \left[\Gamma_i\left(\frac{p_i + \Pi_i}{\rho_i}\right)\right]^{1/2}.
	\label{speedsound}
\end{equation}
These relations reproduce the behaviour of the gas, assumed to be diatomic and perfect, by taking $\Pi_2=0$ and $\Gamma_2=\gamma = 1.4$, with $\gamma$ the ratio of the specific heats. For the liquid, the choice $\Gamma_1=5.5$ and $\Pi_1= 492.115$ MPa is empirically found to reproduce the properties of water \citep{25}.

To non-dimensionalize the above governing equations, we use the initial pressure difference $\Delta p_0 = P_{b,0} - P_\infty$, the initial radius of the bubble $R_0$ and the liquid density $\rho_1$. Hence, we find an inertial timescale $\tau = R_0\left(\rho_1/\Delta p_0\right)^{1/2}$, and a  characteristic velocity $U = R_0/\tau = \left(\Delta p_0/\rho_1\right)^{1/2}$. Times will therefore be multiples of $\tau$ and velocities multiples of $U$. Using these scales, we obtain the following dimensionless groups:
\refstepcounter{equation}
$$
\Rey = \frac{R_0\left(\rho_1\Delta p_0\right)^{1/2}}{\mu_1}, \quad
We = \frac{\Delta p_0R_0}{\sigma}, \quad
Ma = \left(\frac{\Delta p_0}{\rho_1c_1^2}\right)^{1/2},
\eqno{(\theequation{\mathit{a},\mathit{b},\mathit{c}})}\label{eq:dimno1}
$$
which are the Reynolds, Weber and Mach numbers, respectively. The additional dimensionless groups of the problem are:
\refstepcounter{equation}
$$
PR = \frac{P_{b,0}}{P_\infty}, \quad
\chi = \frac{H}{R_0},
\eqno{(\theequation{\mathit{a},\mathit{b}})}\label{eq:dimno2}
$$
 which are the pressure ratio (or compression ratio), and the dimensionless initial distance of the bubble to the fluids' interface, respectively. 

\subsection{Numerical scheme}
We use a finite-volume compressible solver for multiphase flows, introduced by \citet{21}. The interface has a sharp representation, being tracked by a geometric Volume-of-Fluid (VoF) method \citep{23}, where the different phases are labelled by a volume fraction $c$, equal to $1$ in fluid 1, to $0$ in fluid 2, and to values in between for the cells containing an interface. This volume fraction obeys the following advection equation:
\begin{equation}
	\frac{\p c}{\p t} + \bnabla\bcdot\left(c\boldsymbol{u}\right) = c\bnabla\bcdot\boldsymbol{u},
	\label{eq:advection}
\end{equation}
which advances the interface due to the local velocity field. The fluids' properties such as density and viscosity are defined as arithmetic means $\{\rho,\mu\}=c\{\rho_1,\mu_1\} + (1-c)\{\rho_2,\mu_2\}$. This scheme has the advantage of being fully conservative due to the simultaneous advection of the conserved quantities (e.g. total energy, density, momentum) with the volume fraction $c$, thus avoiding any numerical diffusion between the two phases during the advection step. Another advantage of the used scheme is that it takes into account viscous and surface tension effects, the latter being modelled as continuum surface forces (CSF) in the cells containing an interface \citep{24}. The chosen platform  for implementation is the free software program \textsc{basilisk} \citep{34}, a successor of \textsc{gerris} \citep{22}, that has a quad/octree discretization allowing both static and adaptive mesh refinement. For a complete overview of the algorithms involved, the reader should refer to the corresponding publications cited in this subsection.

Boundary Element Methods (BEM), which assume an inviscid and an incompressible ambient liquid, are often used for the simulation of bubble phenomena. Some of these methods account for the liquid's weak compressibility, at low range of Mach numbers \citep{41,42}. The advantage of the present scheme is that it includes viscosity, allowing to study its effect on the dynamics. Moreover, it is an all-Mach formulation, taking into account the liquid's compressibility  and allowing the capture of travelling waves that can sometimes be crucial to the physical phenomena. In addition, VoF methods naturally handle the breakup of the interfaces (e.g. aspherical collapse of a bubble), while in BEM, special care is needed to reconnect the ruptured interfaces. We should note that these advantages come at a cost in terms of speed and computational power, making the DNS approach much more demanding than the BEM.

\section{Phenomenology}\label{sec:crown}
\begin{figure}
	\centerline{\includegraphics[width=\textwidth]{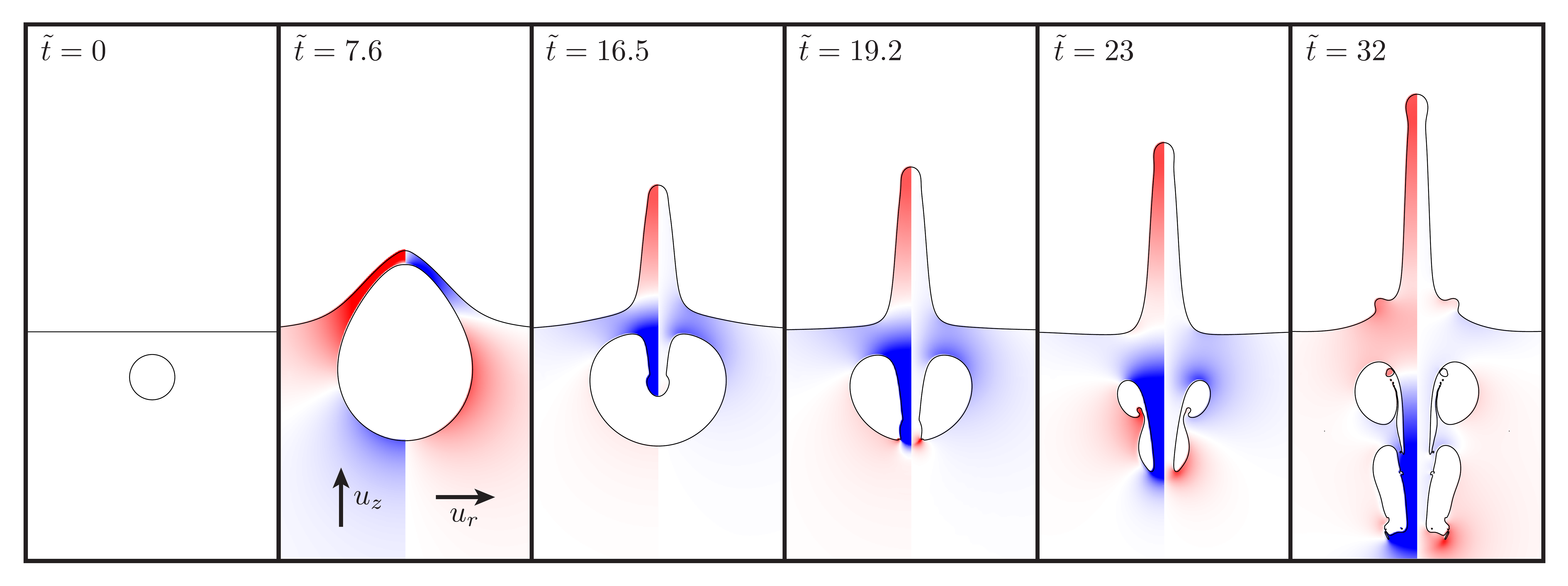}}
	\caption{A numerical sequence of events due to a pressurized bubble at the vicinity of the free surface. For each instant, the respective snapshot is divided into two panels: on the left, the magnitude of the vertical velocity in the liquid is shown as colour code, whereas on the right, the magnitude of the radial velocity is presented. The red colour means positive velocity, in the direction of the defined arrows, while blue stands for negative velocities, and stagnation points are shown in white. For this simulation, $\Rey\rightarrow\infty$, $We = 1000$, $Ma = 0.05$, $PR = 20$, and $\chi = 2$.}
	\label{fig:numseq}
\end{figure}

\subsection{Asymmetric bubble collapse and jet formation}
If potential instabilities of the spherical shape do not occur or can be neglected, a bubble in an infinite liquid domain, subject to a pressure difference with its surrounding liquid, collapses in a spherically symmetric fashion. During this collapse, the internal pressure increases and reaches its peak when the bubble is at its minimum volume. At this specific instant, the internal energy stored by the bubble also reaches a peak, and the high internal pressure generates a compression wave in the liquid which can evolve into a shock or be a shock already at the bubble interface if the collapse is  sufficiently violent, i.e., if the initial pressure difference is large enough \citep{46,43}. In LIFT, however, the bubble is not in an infinite liquid domain, but close to an air-liquid interface, leading to breakage of the spherical symmetry, and to an only axially symmetric collapsing bubble, resulting in an upward jet. 
Past work on LIFT, both experimental \citep{duocastella2010film,3}  and numerical \citep{19,bempedelis2021numerical}, shows that the eventual crown formation after the central jet had formed, is correlated with the second expansion of the bubble and the induced outward flow. However, since the shock wave is emitted at the same time, it is difficult to disentangle these two effects and conclude on the origin of crown formation. Therefore, to settle this question, we recur to numerical simulations, where one can easily vary the control parameters over a large span, in particular including low pressure ratios ($PR$) so as to weaken the compression shock and check whether a crown still forms.

Typically, in an experiment, the initial pressure inside the bubble is unknown. However, it is a key parameter in any numerical simulation, and a good prediction of its value is a prerequisite for any decent comparison between experiments and numerics. A good prediction of $P_{b,0}$ is one that reproduces the experimental inflation of the bubble, or in other words one that is enough for the bubble to reach its experimental maximum equivalent radius $R_m$, given its initial radius $R_0$. If the bubble was in an infinite medium, $P_{b,0}$ would be well predicted by the Rayleigh-Plesset equation or by one of its weakly compressible variants. Although the spherical symmetry is broken in our current setup, the use of such equations is still useful and greatly limits the number of trial and error iterations needed for a good prediction of $P_{b,0}$ \citep{18}. The latter authors also point out the great computational cost and difficulties associated with simulating the experimental parameters. In their simulation, given an initial radius of $0.1\ mm$, a pressure ratio of $2180$ was found to reproduce the experimental maximum equivalent radius $5.2\ mm$. Note that we do not aim to present one-to-one comparisons with the experiments where $PR \sim O(10^3)$. This allows us to make further compromises with the experimental parameters such as assuming a bigger $R_0$, and thus pressure ratios  that are orders of magnitude smaller, which were found sufficient for our current purposes. This renders the simulations more manageable. Furthermore, for higher pressure ratios, bursting behaviour was observed at the interface, such as in \citet{19}.

Figure \ref{fig:numseq} shows a typical sequence of events produced by our numerical simulations. The snapshots show a very good qualitative agreement with their respective experimental counterparts in figure \ref{fig:expseq}. At $\tilde{t} = t/\tau = 0$, a spherical bubble is initialized below the free surface. In experiments, when the laser is focused, plasma forms, emitting a shock wave \citep{36}. The latter can be seen in the first snapshot of figure \ref{fig:expseq}. The emitted shock wave hits the flat interface, and a rarefaction wave is reflected, carrying negative pressure. If gaseous impurities exist between the free surface and the laser-induced bubble, and if the reflected pressure is lower than the liquid's vapour pressure, the cavitation of a cloud of tiny bubbles can occur \citep{44,prl-ando-2012}. The bubbles in this cloud can burst at the free surface \citep{4,5,6}, ejecting small droplets which would ruin the print quality in LIFT \citep{17}. This explains the small droplets that can be seen at the tip of the central jet in the subsequent snapshots of figure \ref{fig:expseq}. In our simulations, we do not attempt to model the small secondary bubbles, nor does our numerical model allow spontaneous cavitation. In all cases, by the time the crown forms, this cloud would be long gone, pushed away by the expanding bubble, and therefore having no effect on the crown formation which is our principal focus.

The pressurized spherical bubble expands but soon loses its sphericity as its upper half gets elongated and deforms the free interface while the bottom half remains relatively hemispherical. So at the end of the expansion phase, $\tilde{t} = 7.6$, the bubble has an egg-like shape, with the tip pointing upwards. At $\tilde{t} = 16.5$, an inner jet develops inside the bubble. An important quantity for the explanation of such a jet, and in the analysis of cavitation bubbles in general, is the Kelvin impulse. For details regarding its derivation and its implications on the translatory motion of the bubble, the reader is referred to \citet{27} and \citet{28}. The predictions of this theory fall in line with the law of Bjerknes: a bubble would move towards a rigid boundary but away from a free surface. In other words, the inner jet that pierces the bubble is directed away from a free surface, as can be seen at $\tilde{t} = 16.5$. Meanwhile, the free surface is rising upwards due to the acquired inertia. This results in a stagnation point associated with a high pressure that further drives both the central and inner jets. At $\tilde{t} = 19.2$, the inner jet impacts the bottom wall of the bubble and breaks it into a toroidal structure. At $\tilde{t} = 23$, we see that the torus further breaks into two toroids, due to shearing instabilities. The bubble is nearing its minimum volume and locally, the free surface slightly moves downwards for it has to fill the void created by the collapsing bubble and its downwards migration. The same behaviour is noticed in experiments as well ($t =130 \ \mu s$ in figure \ref{fig:expseq}). At $\tilde{t} = 32$, we observe the formation of a crown, and the onset of a Rayleigh-Plateau instability at the tip of the central jet, that will eventually lead to pinch-off of a droplet.

\subsection{Crown formation}
\begin{figure}
	\centerline{\includegraphics[width=\textwidth]{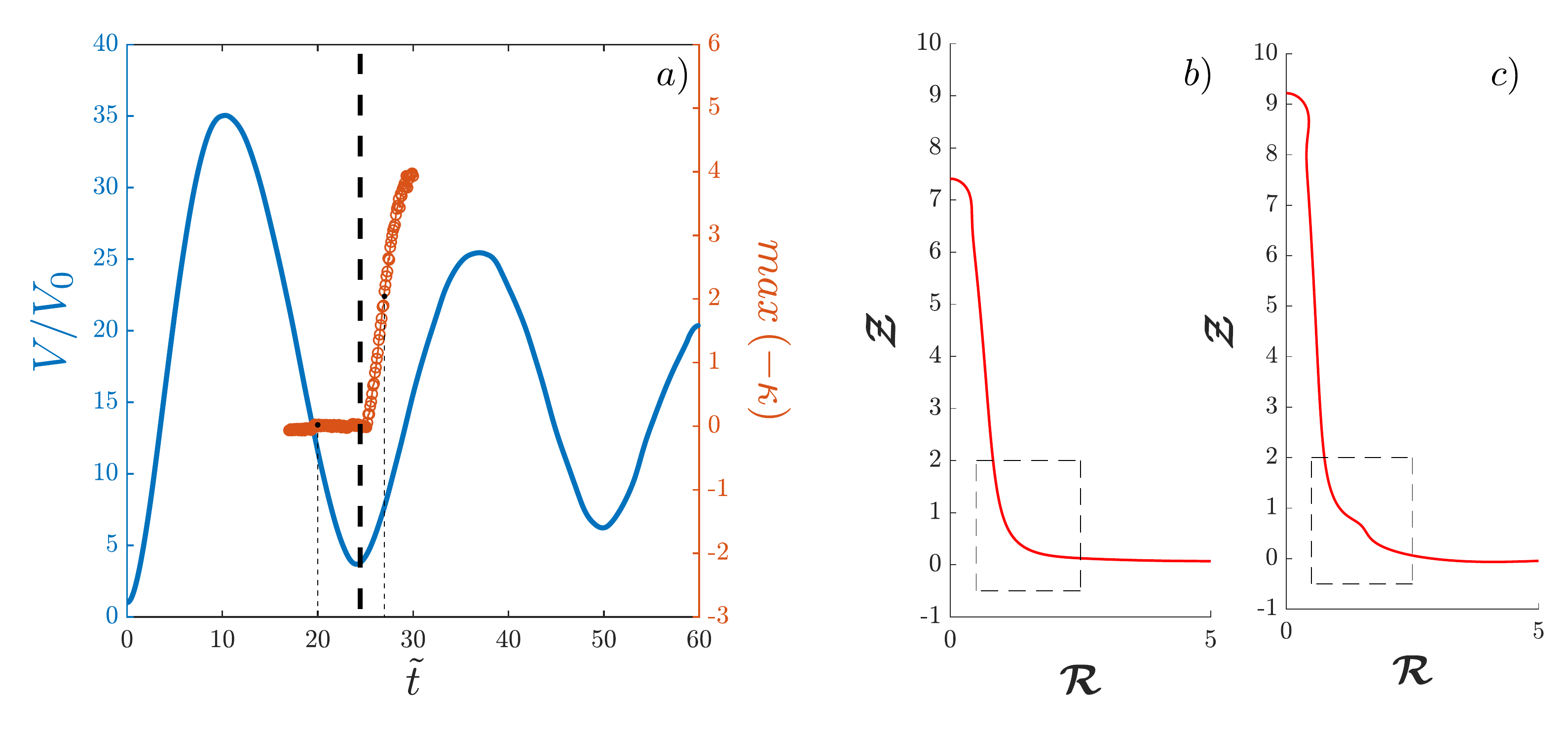}}
	\caption{(a) Relative bubble volume $V/V_0$ and in-plane curvature $\kappa$ as function of time. The blue curve shows the bubble volume. The thick dashed line marks the instant of crown formation, where the value of $max(-\kappa)$,  the orange curve, suddenly increases. (b) Free surface at $\tilde{t} = 20$. The dashed black box shows the location where the curvature $\kappa$ is recorded, from which $max(-\kappa)$ is determined. (c) Free surface at $\tilde{t} = 27$, when the crown has formed. The two instants, $\tilde{t} = 20$ and $\tilde{t} = 27$ of (b) and (c), are marked with thin dashed lines in (a). For this simulation, $\Rey = 2000$, $We = 1000$, $Ma = 0.05$, $PR = 20$, and $\chi = 2$.}
	\label{fig:curvature-volume}
\end{figure}

We now focus on the details of the crown formation. To that end we plot, in figure \ref{fig:curvature-volume}a, a typical example of the normalized volume of the bubble $V/V_0$ as function of time $\tilde{t} = t/\tau$. Figure \ref{fig:curvature-volume}b shows the free surface at $\tilde{t} = 20$, when no crown has formed yet. We thus focus on a region delimited by the black dashed box and record the maximum of $-\kappa$, with $\kappa(r,z) = f_{rr}/(1+f_r^2)^{3/2}$ being the in-plane curvature of the interface $z = f(r)$, positive in the way the latter is defined. Therefore, $max(-\kappa)$ is zero, owing to the virtually flat lines at both ends of the curved interface, and rapidly grows upon the emergence of an inflection point, i.e. upon the formation of the crown. Such a case is illustrated in figure \ref{fig:curvature-volume}c at $\tilde{t} = 27$. We therefore track this quantity in the region of interest, for the time interval in which the crown remains bounded by the black dashed box, and plot it in figure \ref{fig:curvature-volume}a along with the bubble volume. Figures \ref{fig:curvature-volume}b and \ref{fig:curvature-volume}c, marking pre and post crown formation states respectively, are extracted at times shown with thin dashed lines in figure \ref{fig:curvature-volume}a. The formation of the crown seems to be highly correlated with the second expansion of the bubble and happens almost at the same time. We will support this observation with more information on the pressure field.

\begin{figure}
	\centerline{\includegraphics[width=\textwidth]{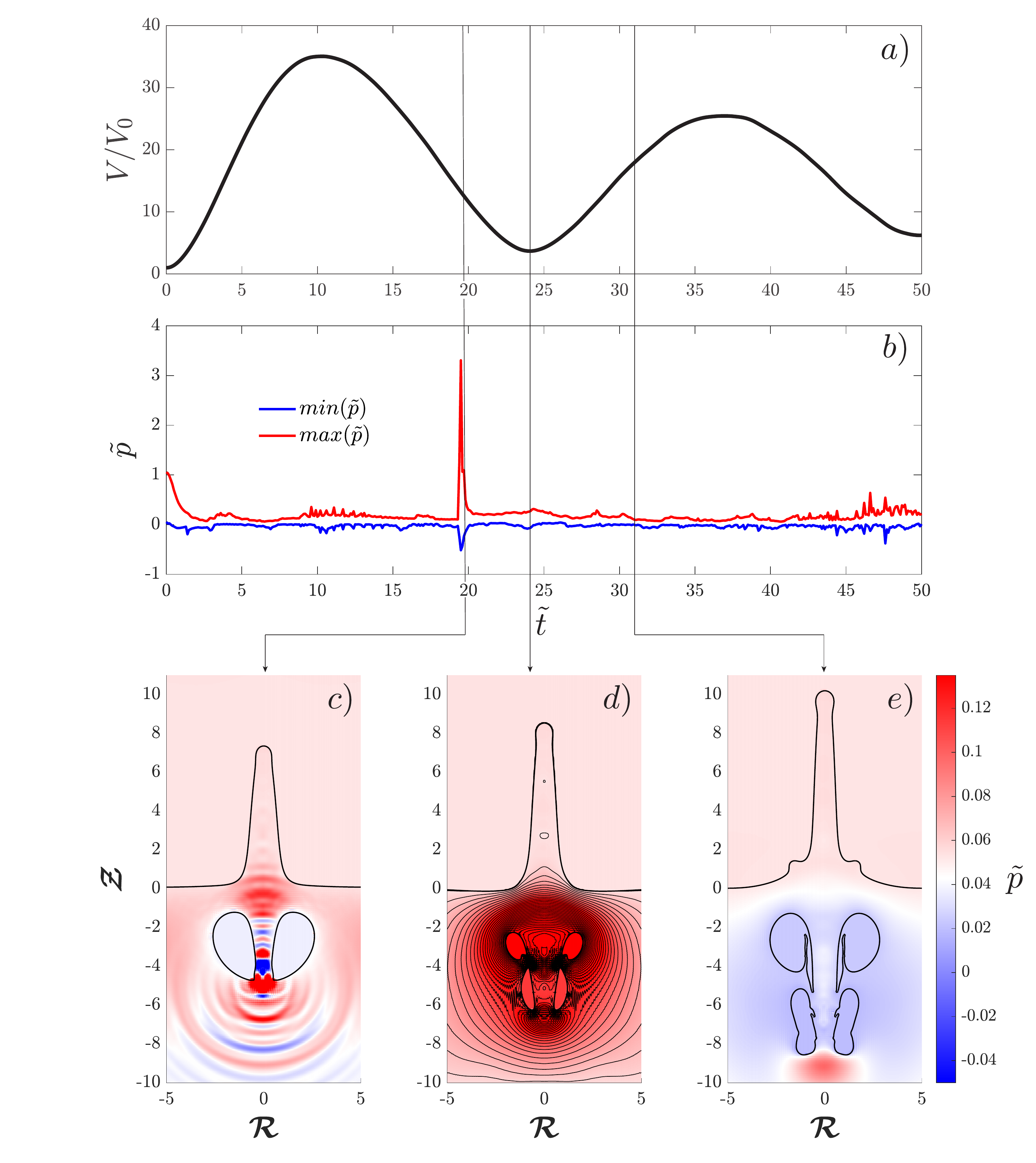}}
	\caption{ (a) Relative bubble volume $V/V_0$ as function of time. (b) Instantaneous pressure signals. The red curve shows the maximum pressure whereas the blue curve records the minimum. (c)(d)(e) Free surfaces and bubble shapes at $\tilde{t} = 19.8$, $\tilde{t} = 24.3$, and $\tilde{t} = 31$, respectively. In (d), pressure isobars are plotted as well. The solid lines with arrows help in marking the respective instants in time. The colour code is for the pressure field. For this simulation, $\Rey = 2000$, $We = 1000$, $Ma = 0.05$, $PR = 20$, and $\chi = 2$.}
	\label{fig:pps}
\end{figure}

In figure \ref{fig:pps}, we plot the life cycle of the same bubble in panel (a) and the recorded peak pressures in panel (b). The strongest shock wave occurs upon the impact of the inner jet and the breakup of the bubble into a toroidal structure (figure \ref{fig:pps}c). In general, during aspherical collapse of a bubble close to a free interface, and next to the shock wave emitted at maximum compression, a water-hammer shock is generated when the inner jet hits the wall of the bubble \citep{30,49,43}. In our simulations, this shock wave is evident, and although part of it is blocked by the bubble itself, it still reaches the free surface and impinges on the curved interface; however, no crown forms (figure \ref{fig:pps}c). At the moment of crown formation, when the bubble is at its minimum volume (figure \ref{fig:pps}d), we do not see the propagation of a shock wave, due to the relatively low chosen pressure ratio $PR$, but rather just an increase in pressure, due to the maximum compression of the bubble. Therefore, the compressibility of the ambient liquid and its ability to sustain and propagate a shock wave seem to have negligible effect in the process of crown formation.

\begin{figure}
	\centerline{\includegraphics[width=\textwidth]{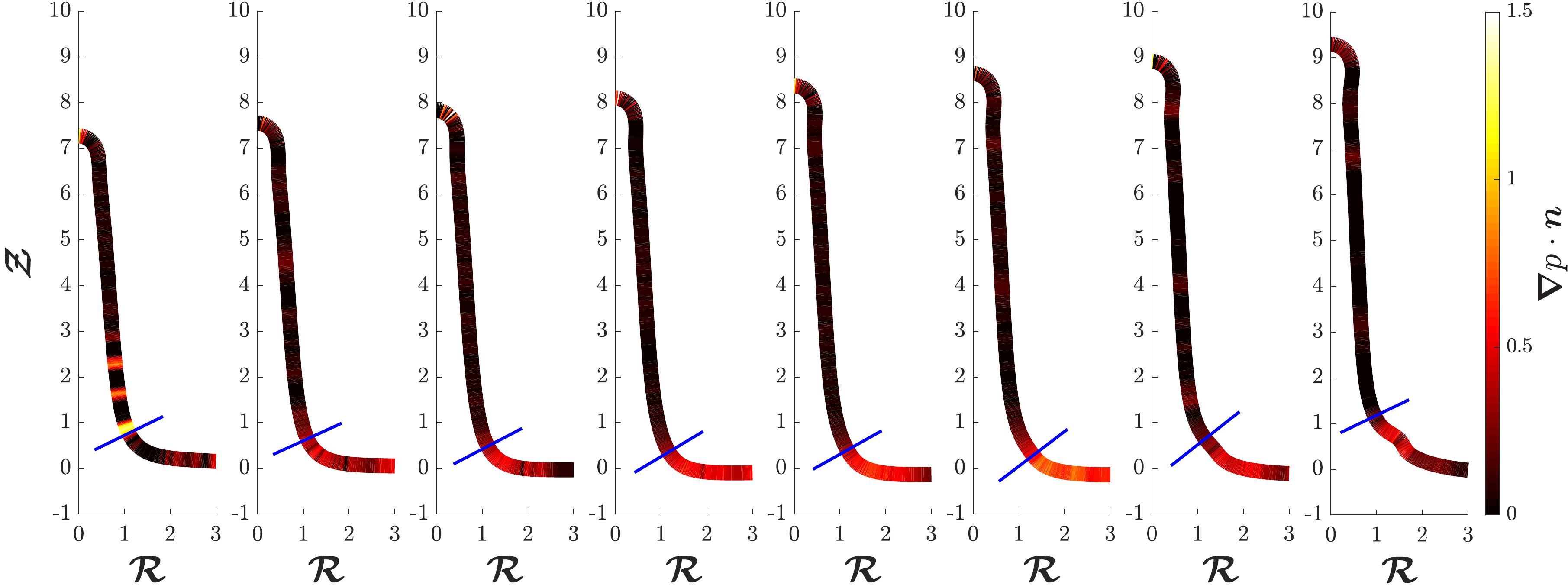}}
	\caption{Shapes of the free surface at $\tilde{t} = 19.5,20.5,21.5,22.5,23.5,24.5,25.5,27$, from left to right, respectively. The interface is coloured by the value of the  normal gradient of pressure. In each snapshot, a blue line indicates the location of zero curvature. For this simulation, $\Rey = 2000$, $We = 1000$, $Ma = 0.05$, $PR = 20$, and $\chi = 2$.}
	\label{fig:gradP}
\end{figure}


A feature we believe is crucial in the formation of the crown is the position of the interface at that moment (figure \ref{fig:pps}d). We previously mentioned that locally, the free surface dips down to fill in the void created by the collapsing bubble. Therefore, when the upper torus expands again, a flow reversal takes place and the surrounding liquid is pushed outwards due to its near incompressibility, resulting in a reversal of the near interface's curvature and direction of motion. Thus, a crown is formed due to the second expansion of the bubble. In addition, the toroids are under high pressure, owing to their maximum compression, i.e. minimum volume. The distorted pressure field around the curved interface, shown by the focusing of isobars in figure \ref{fig:pps}d, contributes to the process by focusing the flow on the curved interface (higher pressure gradient normal to the region of high curvature), but is not sufficient per se to drive the crown. It thus helps in picking the curved interface as the preferential location of crown appearance. Figure \ref{fig:pps}e shows the pressure field at $\tilde{t} = 29$ where the crown has already formed. It must be stated that the other waves that we observe at later stages of the bubble life are due to the collapse of smaller bubbles. Movies of the simulation depicting both the pressure and the velocity fields can be found in the supplementary materials.

Figure \ref{fig:gradP} shows snapshots of the free surface at times $\tilde{t}\in[19.5-27]$, where the interface is coloured by the value of the gradient of pressure in the normal direction. The blue line in each of the snapshots indicates the location of zero curvature.  In 3D, the curvature comprises two parts: an in-plane curvature plotted in figure \ref{fig:curvature-volume}a, and an out-of plane curvature  $\kappa = 1/R$. If one looks at the central jet in the snapshots, its in-plane curvature is null, depicted by the straight line, whereas its out-of plane curvature is convex and positive. The central jet then connects to the flat free surface ($z = 0$), with a concave negative in-plane curvature. Therefore, an annulus exists where the two curvatures cancel out, and where, counter-intuitively, surface tension does not act, as can be inferred from equation \ref{eq:momentum}.
At $\tilde{t} = 19.5$, we see alternating gradients of pressure at distinct patches of the interface. This is due to the propagating water hammer shock, illustrated in figure \ref{fig:pps}c. As the bubble collapses, its internal pressure increases, and so does the gradient of pressure normal to the interface, which peaks around $\tilde{t}  =24.5$ when the bubble reaches its minimum volume. The crown seems to initiate from the zero-curvature annulus which is then shifted upwards. This might be the reason as to why one observes two in-plane micro jets instead of an axisymmetric crown in \citet{20} and \citet{15}, since the axisymmetry of such an annulus is broken in that configuration.

\begin{figure}
	\centerline{\includegraphics[width=\textwidth]{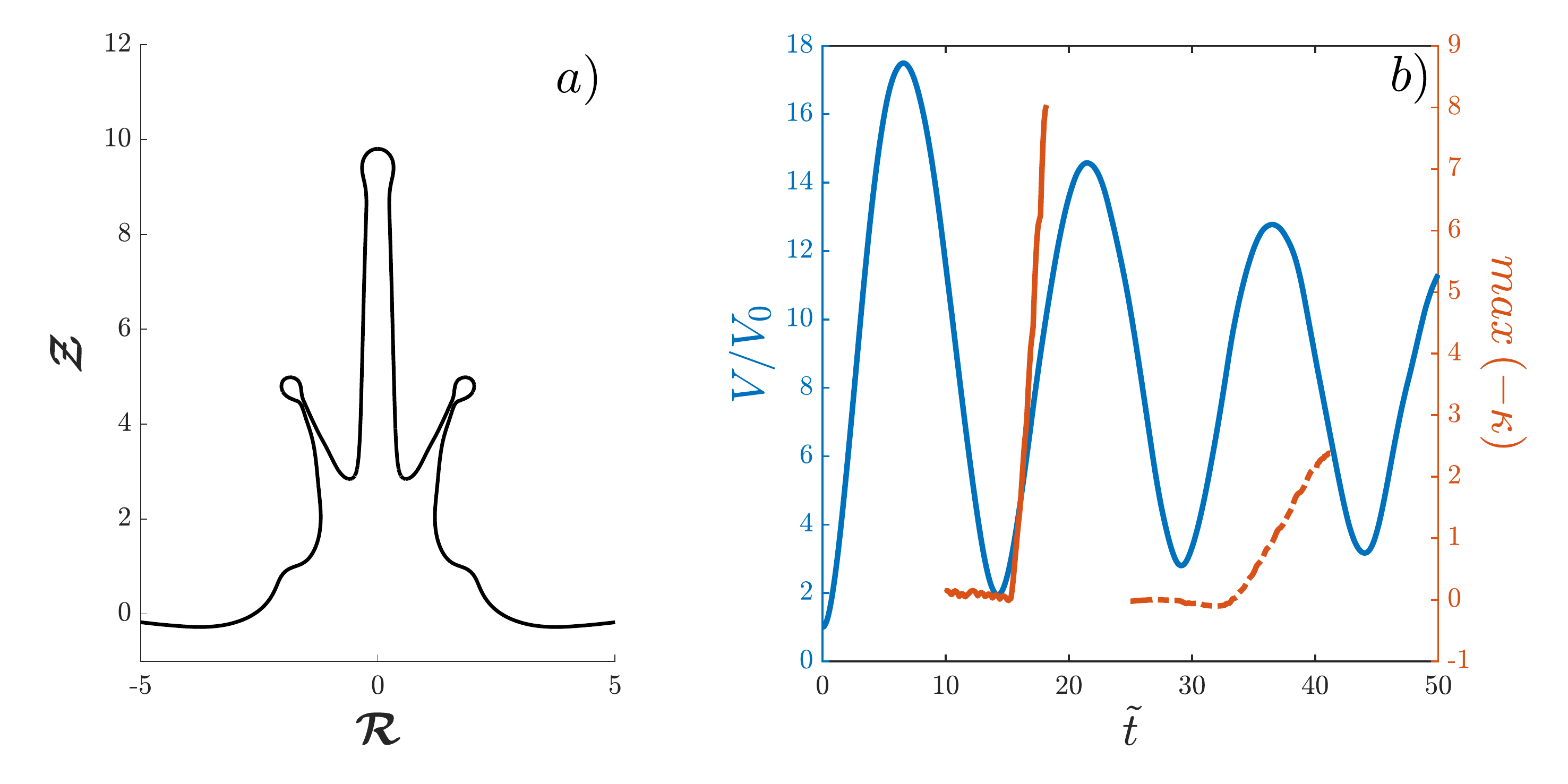}}
	\caption{ (a) Free surface at $\tilde{t} = 40$. (b) The blue curve shows the bubble volume as function of time. The solid orange curve records the value of $max(-\kappa)$ in the region where the crown is formed. The dashed orange line records it in the region where the ``secondary crown" is formed. For this simulation, $\Rey\rightarrow\infty$, $We = 4000$, $Ma = 0.05$, $PR = 10$, and $\chi = 2$.}
	\label{fig:secondarycrown}
\end{figure}

At high $We$, when inertia is much stronger than the capillary forces, we also observe the onset of a ``secondary crown". An example is shown in figure \ref{fig:secondarycrown}a. Figure \ref{fig:secondarycrown}b tracks the topological changes in the respective regions of interest, similarly to the dashed boxes in figures \ref{fig:curvature-volume}b and \ref{fig:curvature-volume}c. The solid orange line corresponds to the primary crown whereas the dashed orange line corresponds to its ``secondary" counterpart. We clearly see that the latter also occurs due to the third expansion of the bubble. However, since the bubble rebounds become weaker with time, these topological features also become less pronounced. For instance, it takes more time to reverse the curvature of the interface and thus the increasing delay between the minima of the blue curve and the associated jumps in curvature. A movie of the simulation is included in the supplementary material.
Also, counter-intuitively, the formation of a ``secondary crown" is weakened by higher pressure ratios, since then the bubble tends to migrate faster downwards and away from the free surface. This is why one rarely observes these topologies in experiments. This presents further evidence as to how and why the rebound of the bubble drives the crown.

\section{Parametric study}\label{sec:param}
\begin{figure}
	\centerline{\includegraphics[width=\textwidth]{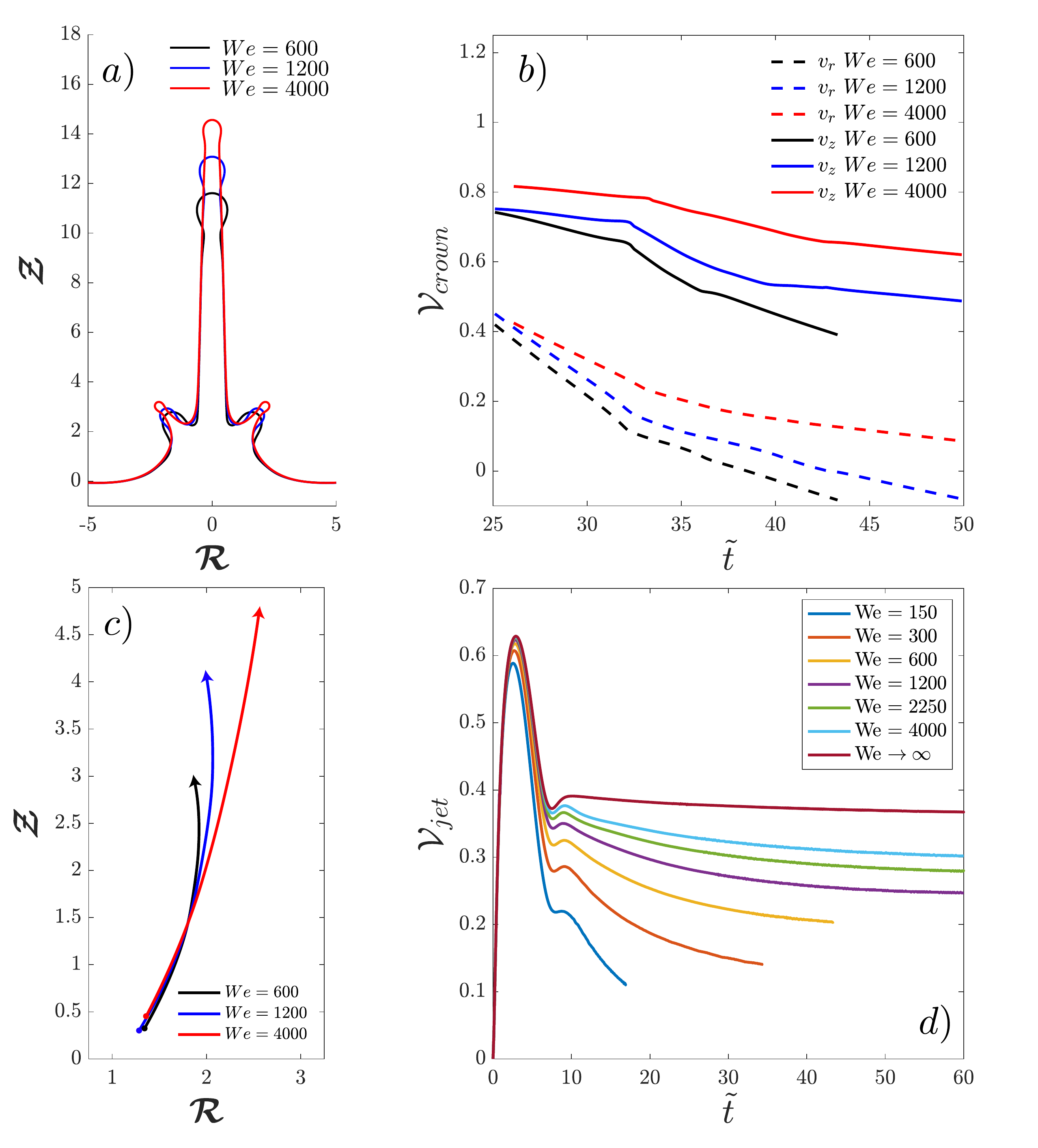}}
	\caption{ (a) Free surface shape at $\tilde{t} = 40$ for different Weber numbers $We$. (b) Instantaneous radial (dashed) and axial (solid) velocities of the crown's tip for different $We$. (c) Trajectories followed by the crown's tip for different $We$. The dots mark the beginning of the time interval whereas the arrows mark its end as well as the direction of motion. (d) Instantaneous central jet velocity measured from its tip, for different $We$. For these simulations, $\Rey\rightarrow\infty$, $Ma = 0.05$, $PR = 20$, and $\chi = 2$.}
	\label{fig:weber}
\end{figure}

The dynamic processes, explained above, depend on several physical parameters. Hence, in this section, we perform a parametric study. Out of the five dimensionless numbers that we defined in section \ref{subsec:eqs}, we varied the Weber number $We$, the Reynolds number $\Rey$, the initial pressure ratio $PR$ and the dimensionless bubble distance to the free surface $\chi$. In varying the Weber number, we study the effect of surface tension. Experimentally speaking, surface tension effects can be changed by changing the liquid or by adding surfactants, provided they instantaneously spread. The viscosity can also be readily changed by mixing water with other fluids (e.g. glycerol), and thus obtaining different Reynolds numbers $\Rey$. For the standard operating parameters of the experiments, $\Rey\sim O(10^4)$, so one can fairly use an inviscid flow approximation \citep{26} if one's aim is to compare one-to-one with the experiments. Furthermore, the pressure ratio $PR$ increases with the laser energy. Finally, one can vary $\chi$ by focusing the laser at different depths. In this parametric study, we perform all simulations for a fixed Mach number $Ma = 0.05$ in the weakly compressible surrounding liquid.

\begin{figure}
	\centerline{\includegraphics[width=\textwidth]{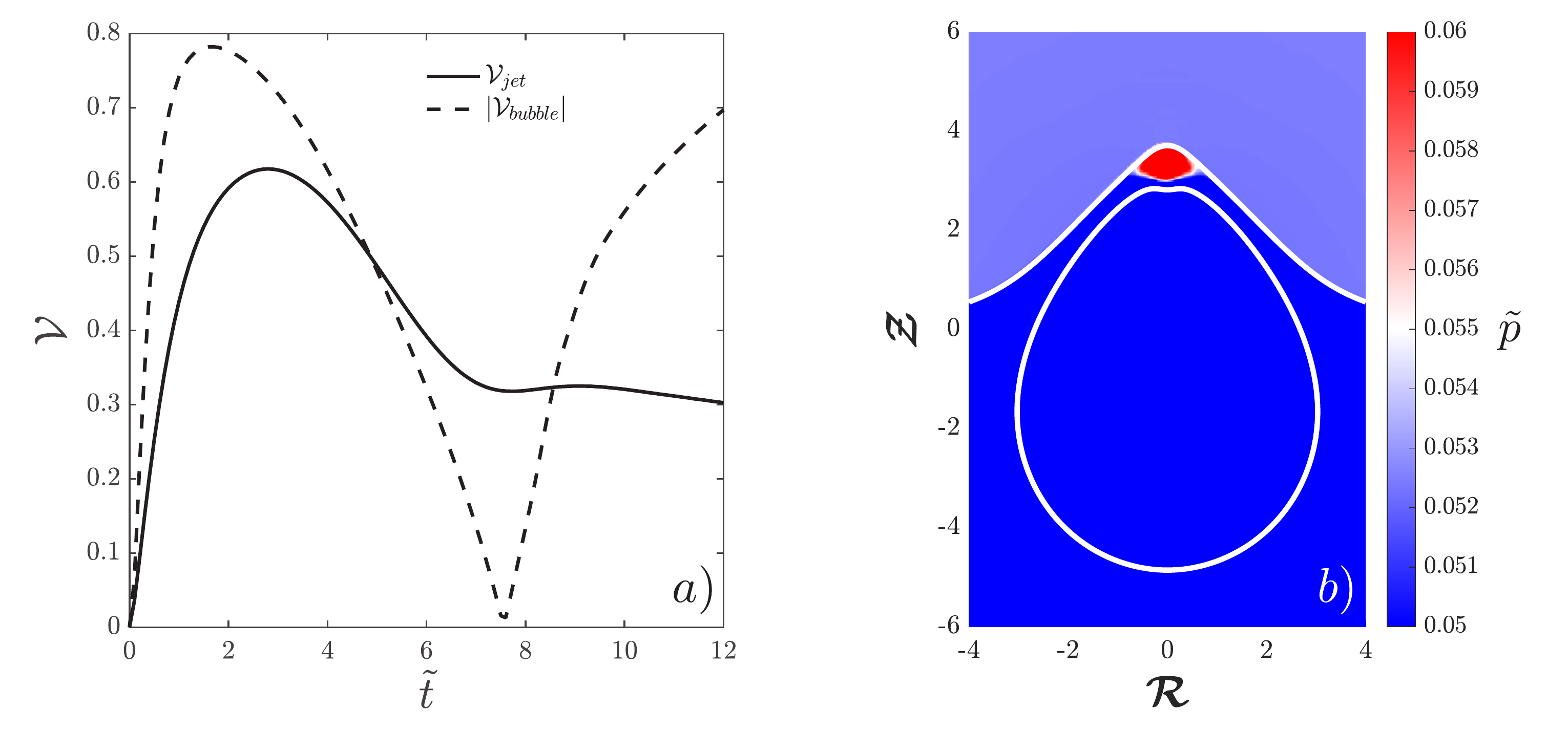}}
	\caption{ (a) The solid line shows the instantaneous jet velocity measured from the motion of its tip. The dashed line shows the velocity of the upper tip of the bubble, in absolute value. (b) Interfaces at $\tilde{t} = 8$. The colour code is for the pressure field. For this simulation, $\Rey\rightarrow\infty$, $We = 600$, $Ma = 0.05$, $PR = 20$, and $\chi = 2$.}
	\label{fig:staticp}
\end{figure}

\subsection{Dependence on the Weber number $We$}
Figure \ref{fig:weber}a shows the effect of the Weber number on the jet and crown formation. The larger the $We$ (smaller surface tension), the faster the central jet. The maximum thickness of the jet does not seem to be affected by surface tension, and, for the different values of $We$, the jets thin similarly as they extend upwards. In addition, as $We$ increases, the crown gets more pronounced, i.e., it increasingly resembles a growing thin jet instead of a bulge of liquid. Figure \ref{fig:weber}b shows the instantaneous radial and axial velocities of the crown tips for different Weber numbers. The larger the $We$, the larger the crown velocity. At small $We$, surface tension forces are strong and tend to reconnect the crown with the central jet via capillary waves. This can be seen by the negative radial velocity at later times. Figure \ref{fig:weber}c depicts the trajectories followed by the crown tips. For relatively low $We$, we see again the crown's tendency to rejoin the central jet (arrows pointing towards the $z$-axis). However, for large $We$, the crown stays separate and eventually pinches off into a detached torus, as is suggested by figure \ref{fig:secondarycrown}a. An axisymmetric model cannot evidently simulate the secondary break-up of the crown into droplets by the Rayleigh-Plateau instability. A full 3D simulation, allowing variations in the out-of-plane curvature, is therefore needed to capture this phenomenon. In contrast, the development of the Rayleigh-Plateau instability on the central jet can be captured with the present axisymmetric simulations. At long times, this instability results in a spherical droplet at the tip of the central jet, the inset of which is shown in figure \ref{fig:weber}a. Figure \ref{fig:weber}d shows the instantaneous dimensionless velocity profiles $\mathcal{V}_{jet} = \dot{z}(t)$ of the free surface, at $r = 0$, for different values of the Weber number, ranging from as low as we could numerically go, to infinity. For small $\tilde{t}$, the free surface accelerates due to the rapid expansion of the bubble beneath it. Surface tension seems to have no effect on this first phase as all the curves for different $We$ collapse. The velocity then reaches a maximum value before deceleration. The higher the $We$, the larger this maximum velocity. The jets eventually reach a constant speed where they rise due to the liquid's acquired inertia. At large $We$, where inertial effects dominate, the jet velocity asymptotically reaches a plateau, i.e. the velocity becomes independent of the capillary effects.

\begin{figure}
	\centerline{\includegraphics[width=\textwidth]{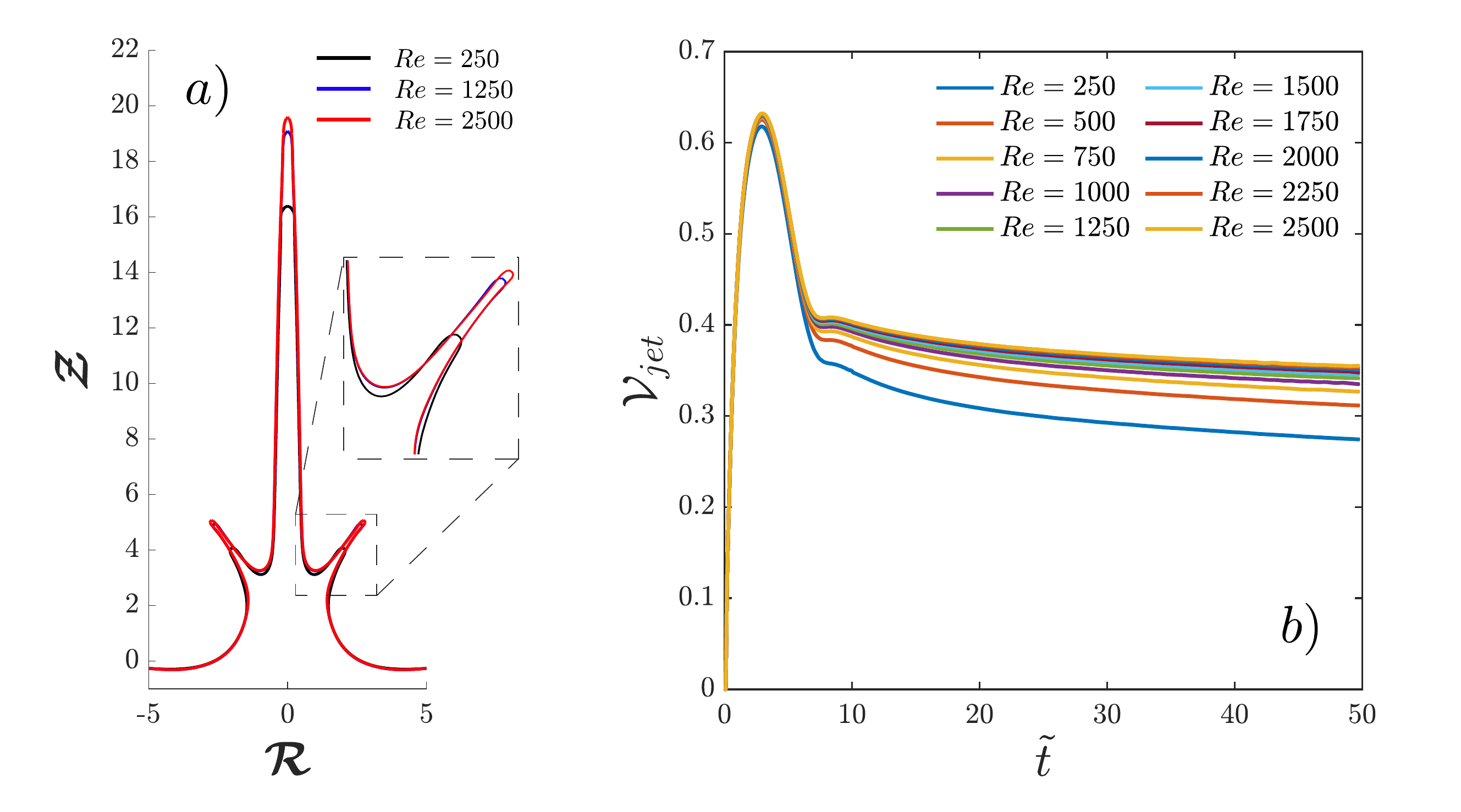}}
	\caption{ (a) Free surface shape at $\tilde{t} = 50$ for three different Reynolds numbers $\Rey$, see legend. (b) Instantaneous central jet velocity measured from its tip, for the different $\Rey$, see legend. For these simulations, $We\rightarrow\infty$, $Ma = 0.05$, $PR = 20$, and $\chi = 2$.}
	\label{fig:Re}
\end{figure}

 The curves in figure \ref{fig:weber}d also feature a second peak (at $\tilde{t}\sim9$ for this set of parameters), but with a lower amplitude. To explain the origin of this kink, in figure \ref{fig:staticp}a we plot the axial tip ($r = 0$) velocities of both the free surface and the upper wall of the bubble. At $\tilde{t}\sim7.6$, the bubble has reached its maximum expansion where its upper wall stops its upward motion ($\mathcal{V}_{bubble} = 0$) and the inner jet starts developing. This velocity (dashed curve) is plotted as absolute value, in order to stress this particular moment in time. One clearly sees that the second surge in the free surface velocity happens at the same time. This was also observed in the boundary integral simulations of \citet{31} and \citet{48}. Physically, as the bubble's tip reverses its motion, a stagnation point develops, associated with a higher static pressure, shown in figure \ref{fig:staticp}b. This pressure repels the fluid axially in both directions, thus accelerating the free surface upwards, and contributing to the formation of both the central and inner jets. Note that all the key features of figure \ref{fig:weber}d, whether the peak velocity or the kink, seem to happen at the same instant in time. We thus conclude that, for the range of Weber numbers considered here,  surface tension has practically no effect on the growth of the bubble which seems to expand in the same way in all the cases that we have considered. 

\subsection{Dependence on the Reynolds number $\Rey$}
Figure \ref{fig:Re}a shows the effect of a varying Reynolds number $\Rey$ on the phenomenon. With increasing $\Rey$, the jet becomes faster and so does the crown. This specific parametric study was done in the limit of an infinite Weber number so as to illustrate the competition between inertia and viscosity, with capillarity playing no role. One notices that indeed the jets no longer break into droplets due to the absence of capillary forces. This has its effect on the crown as well which in time remains a distinct topology, separated from the central jet. This is to be contrasted with figure \ref{fig:weber}a for a case of high surface tension (e.g. $We = 600$) where capillary forces eventually flatten the regions of high curvature and merge the crown with the jet. Similar to figure \ref{fig:weber}a, a varying $\Rey$ does not affect the maximum thicknesses of the jets. Figure \ref{fig:Re}b has the same traits as figure \ref{fig:weber}b. For low values of $\Rey$, viscous forces increasingly act against the liquid inertia, thus leading to lower jet velocities. The curves tend to converge asymptotically with increasing $\Rey$ to the inviscid limit $\Rey\rightarrow\infty$. Note that the respective kinks in the curves also occur at the same time, meaning that viscosity has a negligible effect on the bubble expansion for the range of $\Rey$ studied here.

\begin{figure}
	\centerline{\includegraphics[width=\textwidth]{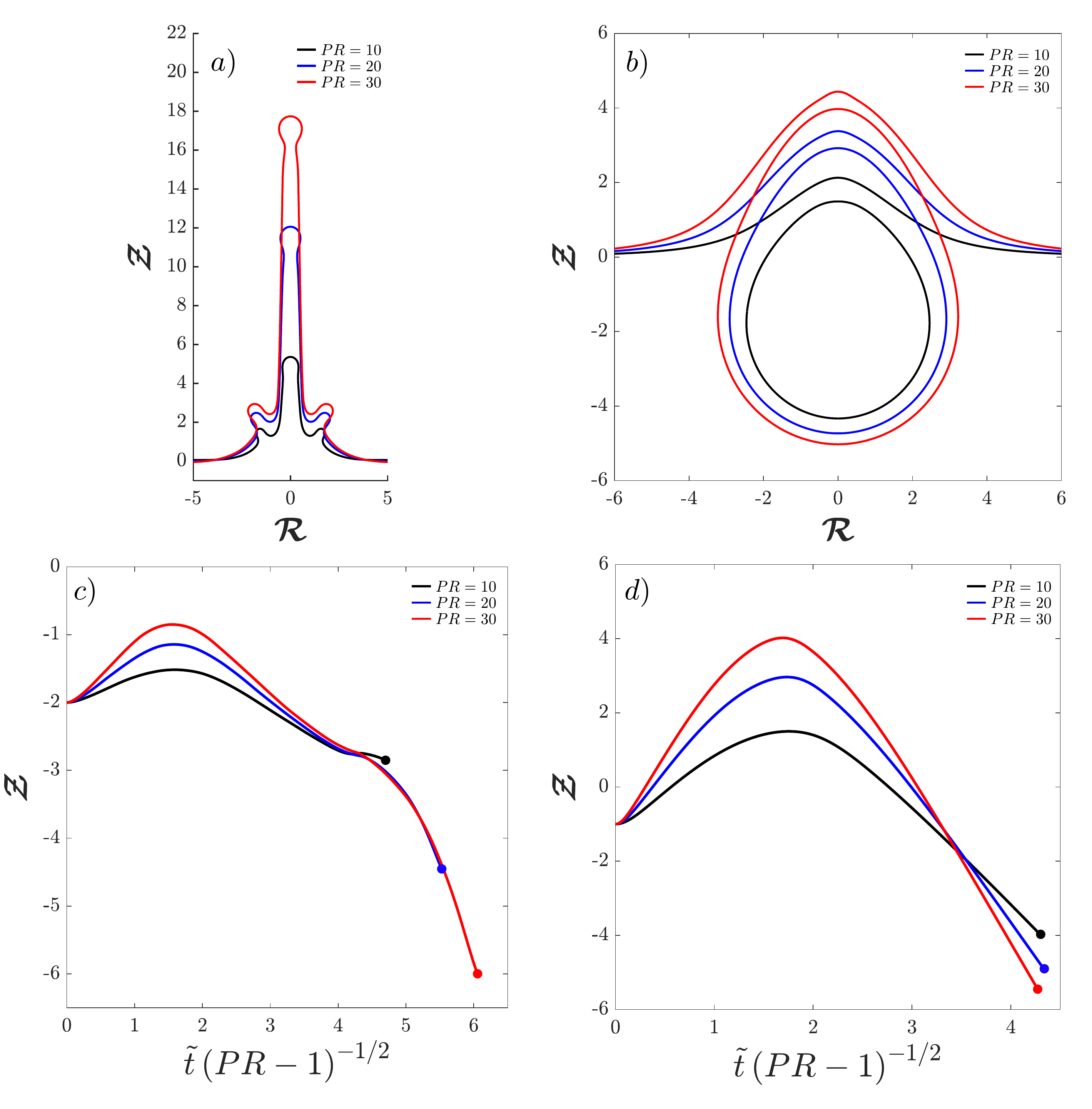}}
	\caption{ (a) The free interface for different values of the pressure ratio $PR$ is shown at analogous instants in time. (b) Egg-shape dependence on the pressure ratio $PR$. (c) Location of the bubble's centroid as a function of time, for the different pressure ratios, up until the respective ends of the first oscillation cycles, where the bubble starts its second expansion. (d) Position of the upper tip of the bubble as a function of time, for the different pressure ratios, up until the respective moments of inner jet impacts leading to the bubble's metamorphosis into a toroidal shape. For these simulations, $\Rey\rightarrow\infty$, $We = 1000$, $Ma = 0.05$, and $\chi = 2$.}
	\label{fig:PR}
\end{figure}

\subsection{Dependence on the pressure ratio $PR$}
Figure \ref{fig:PR}a shows the effect of the pressure ratio on the dynamics. In contrast to the Weber and Reynolds numbers, increasing $PR$ leads to thicker jets. One must think of $PR$ as the amount of energy initially stored inside the bubble. Hence, the larger the $PR$, the  larger this energy, and the more the bubble is going to expand. Given the same bubble position (i.e. the same value of $\chi$), this leads to a more pronounced egg-shape, elongated further in the direction of the free surface (figure \ref{fig:PR}b). Therefore, when the collapse phase begins, the central jet is formed as previously explained, and due to the altered initial deformation of the free surface, its thickness changes. Again, it is the first expansion phase of the bubble that dictates the shape of the central spike. In addition, we see that as PR increases, the central jet becomes faster.

An important quantity in the analysis of cavitation bubbles is the Kelvin impulse defined as:
\begin{equation}
	I = \rho\int_S\phi\boldsymbol{n}dS,
\end{equation}
where $S$ is the surface of the bubble, $\phi$ is the velocity potential and $\boldsymbol{n}$ is the outward normal to the fluid (i.e. into the bubble). This impulse is a measure of the liquid's linear momentum. Considering that the bubble has a source-like behaviour, and by ensuring the global conservation of linear momentum in a control volume encompassing the free surface, the value of $I$ can be found. The Kelvin impulse describes the translatory motion of the bubble in a semi-infinite fluid, and its predictions fall in line with the law of Bjerknes, stating that the oscillating bubble migrates away from the free surface. For details regarding the derivation of the Kelvin impulse, the reader is referred to \citet{27}, \citet{44}, and \citet{28}. \citet{16} derived the expression of the Kelvin impulse for the asymmetric growth and collapse of a bubble in the vicinity of a free surface,
\begin{equation}
	I = 0.934R_m^3\sqrt{\rho_l(P_\infty - P_v)}\left(\frac{H}{R_m}\right)^{-2}\boldsymbol{n}\simeq0.934R_m^3\sqrt{\rho_lP_\infty}\left(\frac{H}{R_m}\right)^{-2}\boldsymbol{n},
	\label{eq:impulse}
\end{equation}
where $R_m$ is the maximum radius achieved by the bubble, and $P_v$ is the vapour pressure. With increasing $PR$, figure \ref{fig:PR}b shows that further inflation is achieved by the bubble, and thus higher values of $R_m$. By inspection of equation \ref{eq:impulse}, one readily sees that $I$ increases with increasing $R_m$, given that all other parameters are kept the same. This means that with increasing $PR$, the bubble travels further downwards, away from the free surface. Figure \ref{fig:PR}c shows the location of the bubble's centroid as a function of time, up until the end of the first oscillation cycle, for the different pressure ratios. During the expansion into the egg-shape, the centroid rises and when the collapse phase begins, the inner jet develops and the bubble migrates downwards, away from the free surface ($z = 0$). The distance travelled by the bubble clearly increases with increasing $PR$. This is why one does not observe ``secondary crowns" with higher pressure ratios, as previously mentioned. Figure \ref{fig:PR}d shows the instantaneous position of the upper tip of the bubble, up until the inner jet impacts the lower wall of the then toroidal bubble. Again, one sees the expansion phase, and then the inner jet that pierces the bubble, with a velocity that increases with increasing $PR$.

\subsection{Dependence on the dimensionless bubble-interface distance $\chi$}
\begin{figure}
	\centerline{\includegraphics[width=\textwidth]{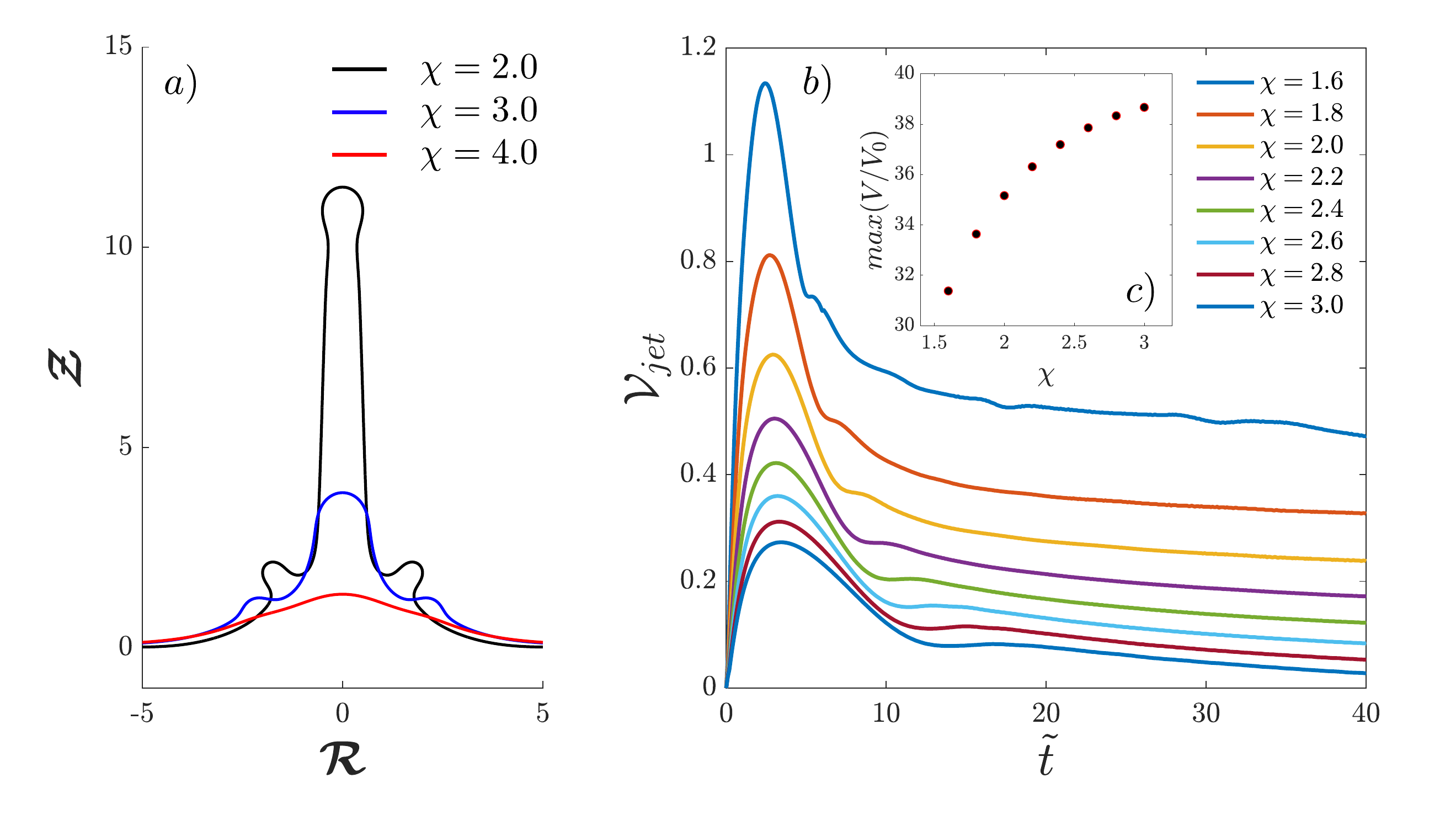}}
	\caption{ (a) Interfaces at $\tilde{t} = 35$ for three different dimensionless bubble-interface distances, $\chi$. (b) Instantaneous jet velocity for different values of $\chi$. (c) The inset shows the maximum volume reached by the expanding bubble as a function of $\chi$. For these simulations, $\Rey\rightarrow\infty$, $We = 1000$, $Ma = 0.05$, and $PR = 20$.}
	\label{fig:H}
\end{figure}

Figure \ref{fig:H}a shows the effect of the dimensionless bubble distance from the free surface, $\chi$, on the dynamics of both jet and crown. As $\chi$ increases, the bubble retains its spherical shape as compared to an elongated egg-shape when the bubble is closer to the free surface. This is crucial to the formation of the central jet and, subsequently, the crown, as previously explained. These features will be suppressed for too large $\chi$, when only a bump occurs on the free surface (e.g., for $\chi = 4.0$). In order to compensate for the lack of proximity to the free surface (too large $\chi$), one can increase the pressure ratio $PR$, and thus achieve a further expansion of the bubble, then again leading to an egg-shape. Figure \ref{fig:H}b shows that $\chi$ seems to be the most sensitive parameter for which the slightest change creates a large velocity difference. All the curves exhibit the same behaviour as in figures \ref{fig:weber}d and \ref{fig:Re}b. As the distance to the free surface gets smaller, the central jet velocity becomes larger. On the other hand, from the inset \ref{fig:H}c, one sees that as $\chi$ increases, the maximum volume achieved by the bubble expansion slightly increases as well, leading to a delay in the formation of the inner jet and thus the appearance of the kink. A bubble expands if its internal pressure $P_b$ is higher than the surrounding pressure $P_\infty$. As it inflates, the bubble pressure $P_b$ decreases until the point where $P_b = P_\infty$. From this point on, the bubble continues its expansion due to the acquired inertia of the surrounding liquid. As the distance from the free surface $\chi$ increases, the bubble is submerged in a larger liquid mass, and therefore inertial effects are stronger, leading to further inflation. At maximum expansion, the bubble pressure $P_b$ is thus higher for smaller values of $\chi$.

\section{Conclusion}\label{sec:conclusion}
In this paper, we performed numerical simulations  of a cavitating bubble close to a free surface. We observe the formation of a central jet protruding upwards while an inner jet pierces the bubble and breaks it into a toroid. We also observe the formation of a crown highly correlated with the second expansion of the bubble. Further examination of the pressure signal leads us to the conclusion that such a topological feature is not the result of shock wave impingement on the curved free surface, but rather, a combination of a pressure distortion over the curved interface, leading to a preferential selection of the location of crown formation, and of a flow reversal, caused by the second expansion of the toroidal bubble that drives this crown. For high Weber numbers, we also observe weaker ``secondary crowns", correlated with the third oscillation cycle of the bubble.

We performed a parametric study, varying the Weber number $We$, the Reynolds number $\Rey$, the pressure ratio $PR$ and the dimensionless bubble distance from the free surface $\chi$. For a given $\chi$, the central jet's thickness seems to be only affected by the pressure ratio that dictates the shape of the expanding bubble, and thus the deformation of the free surface. All the studied parameters have a direct effect on the velocity of both the central jet and the crown. Varying $\chi$ seems to have the most effect on the formed topologies (jet and crown) which can be reduced merely to bumps provided that the bubble is far enough from the free surface.

This study considered Newtonian fluids only. However, further numerical studies could provide insight into laser-induced forward transfer (LIFT) in complex fluids, i.e. viscoplastic/viscoelastic fluids, similar to what have been experimentally studied by \citet{unger2011time}, \citet{ turkoz2018impulsively}, and \citet{3}. This is most useful for engineering and biomedical applications, and could allow us to shed more light on cavitation in soft matter. Another route of relevant extension of this work is towards liquid pools with finite depths, where the bubble's interaction with the bottom wall becomes important.

\section*{Acknowledgements}
The authors would like to thank Daniel Fuster for stimulating discussions, and Dave Kemper for providing the experimental images to this manuscript. The research leading to these results has received funding from the European Union’s Horizon 2020 Research and Innovation programme under the Marie Skłodowska-Curie Grant Agreement No 813766. This work was carried out on the national e-infrastructure of SURFsara, a subsidiary of SURF cooperation, the collaborative ICT organization for Dutch education and research.

\section*{Declaration of Interests}
The authors report no conflict of interest.

\appendix
\section{Numerical details}\label{appA}
\begin{figure}
	\centerline{\includegraphics[width=\textwidth]{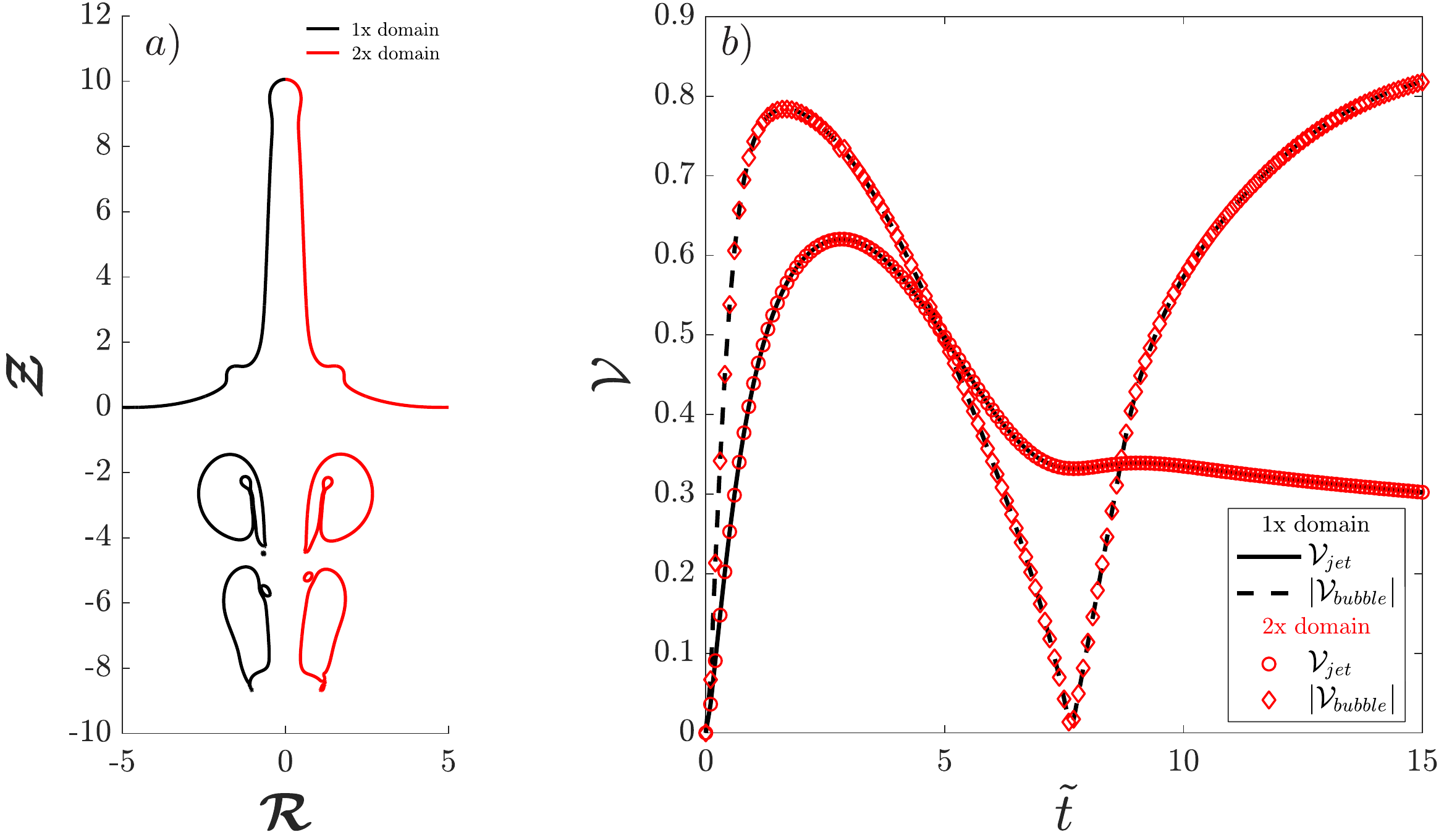}}
	\caption{(a) Shape of the interface at $\tilde{t} = 30$ for two different domain sizes. (b) Instantaneous velocities of the free surface jet, and of the upper tip of the bubble, the latter shown in absolute value, for two different domain sizes. For both simulations, $\Rey = 2000$, $We = 1000$, $Ma = 0.05$, $PR = 20$, and $\chi = 2$.}
	\label{fig:domain}
\end{figure}
In this appendix, some of the simulations' numerical details will be presented. A static mesh is employed with maximum refinement around the interfaces and the regions of interest. The mesh is then progressively coarsened as the boundaries are approached. Since Non Reflective Boundary Conditions (NRBC) are not implemented in the current version of the solver, this coarsening serves as a numerical dissipation of the emitted/reflected  shock/pressure waves that are mostly generated at locations where the grid  is at maximum refinement. Therefore, the highly resolved and localised fronts (pressure peaks) of the waves get flattened via interpolation on the coarser grid. It is imperative that this coarsening be made progressively; otherwise, the waves will be spuriously reflected and will contaminate the regions of interest. In addition, and as mentioned in subsection \ref{subsec:set}, the size of the numerical domain is chosen large enough to mitigate the lack of NRBC. Figure \ref{fig:domain} shows a couple of simulations depicting the effect of the numerical domain size where``1x domain" refers to the actual domain size employed for this paper's simulations, and ``2x domain" twice that size. Figure \ref{fig:domain}a shows the shape of the interface at $\tilde{t} = 30$. Both simulations are virtually identical, especially the free surface topology, i.e. the central jet and the crown. There are slight discrepancies at the level of the toroidal bubble which clearly do not interfere with the dynamics of the free surface. Figure \ref{fig:domain}b shows  the instantaneous velocity $\dot{z}(t)$ of both the free surface and the upper tip of the bubble recorded at $r = 0$. The curves are similar to those of figure \ref{fig:staticp}a, enclosing the same physics, and identical for both domain sizes. This proves that the currently used size, i.e. 50 times the initial radius of the bubble, added to the progressive mesh coarsening strategy, is safe enough to model the physical process at hand.

The maximum resolution used is $\sim 40$ cells per initial bubble radius, and then progressively coarsened as previously discussed. The timestep is governed by an acoustic CFL condition based on the maximum speed of sound in the least compressible fluid, i.e. the liquid,
\begin{equation}
	C = max\{c(r,z)\}\frac{\Delta t}{\Delta x} = max\Bigg\{\left[\Gamma_1\left(\frac{p_1(r,z) + \Pi_1}{\rho_1}\right)\right]^{1/2}\Bigg\}\frac{\Delta t}{\Delta x} = 0.5.
\end{equation}
This condition guarantees that the pressure waves, that travel at the speed of sound, are well captured.

\bibliographystyle{jfm}
\bibliography{lift}

\end{document}